\documentclass[conference]{IEEEtran}
\IEEEoverridecommandlockouts

\usepackage{cite}
\usepackage{amsmath,amssymb,amsfonts}
\usepackage{algorithmic}
\usepackage{algorithm}
\usepackage{graphicx}
\usepackage{stmaryrd}
\usepackage{xcolor}
\usepackage{array}
\usepackage{commath}
\usepackage{sidecap}
\usepackage{stfloats}
\usepackage{tabularx, boldline}
\usepackage{rotating,booktabs,multirow}
\usepackage{mathtools}
\usepackage{flexisym}
\usepackage{breqn}
\usepackage{url}
\usepackage{adjustbox}
\usepackage{makecell}

\usepackage{textcomp}
\usepackage{amsmath}
\usepackage{txfonts}
\usepackage{graphicx}
\usepackage{acronym}
\usepackage{tabularx, boldline}
\usepackage{rotating,booktabs,multirow}
\usepackage{enumerate}

\def\R{\mathbb{R}} 

\begin{document}

\title{Privacy-Preserving Adversarial Network (PPAN) for Continuous non-Gaussian Attributes \\
}

\author{
\IEEEauthorblockN{Mohammadhadi~Shateri}
\IEEEauthorblockA{Department of Electrical and Computer Engineering\\
                    McGill University, Montreal, Canada\\
                    Email: mohammadhadi.shateri@mail.mcgill.ca}\\
\and
\IEEEauthorblockN{Fabrice~Labeau}
\IEEEauthorblockA{Department of Electrical and Computer Engineering\\
                    McGill University, Montreal, Canada\\
                    Email: fabrice.labeau@mcgill.ca}

}

\maketitle

\thispagestyle{plain}
\pagestyle{plain}

\begin{abstract}
    A privacy-preserving adversarial network (PPAN) was recently proposed as an information-theoretical framework to address the issue of privacy in data sharing. The main idea of this model was using mutual information as the privacy measure and adversarial training of two deep neural networks, one as the mechanism and another as the adversary. The performance of the PPAN model for the discrete synthetic data, MNIST handwritten digits, and continuous Gaussian data was evaluated compared to the analytically optimal trade-off. In this study, we evaluate the PPAN model for continuous non-Gaussian data where lower and upper bounds of the privacy-preserving problem are used. These bounds include the Kraskov (KSG) estimation of entropy and mutual information that is based on k-th nearest neighbor. In addition to the synthetic data sets, a practical case for hiding the actual electricity consumption from smart meter readings is examined. The results show that for continuous non-Gaussian data, the PPAN model performs within the determined optimal ranges and close to the lower bound.
\end{abstract}
\begin{IEEEkeywords}
Privacy preserving adversarial network, Adversarial training, Mutual information, Continuous non-Gaussian attributes, Private attribute.
\end{IEEEkeywords}

\section{Introduction}

In the era of data mining, the explosion of data collection is enabling the researchers to increasingly learn and develop sophisticated models on real-life data sets. There are thousands of research works in many areas including medicine, education, electrical power, or business in which data mining tools are applied to large published data sets. Unfortunately, these data sets may include private information about individuals. Therefore, data collection and curation organizations should sanitize the data before releasing/sharing it. The privacy-preserving data sharing models aim at generating a released data $Z$ from an observation $W$ which is a sanitized version of useful/public attribute $Y$ and at the same time minimizes the possibility of inferring about the sensitive/private attribute $X$. 

An information-theoretical approach based on distorting useful data was proposed in \cite{rebollo2010t,sankar2013utility,huang2018generative,tripathy2019privacy,shateri2019deep} where using the Kullback-Leibler(KL) divergence, the privacy risk was quantified as the mutual information between private attribute $X$ and released/shared data $Z$. In \cite{rebollo2010t} for the Gaussian attributes, they used a numerical procedure (a modification of the steepest descent algorithm) to solve this privacy-distortion trade-off. The results were compared with the quadratic-Gaussian lower and upper bounds of mutual information. However, it was not discussed how including the private attribute in observation samples \textemdash $W=\left(X,Y\right)$ compared with $W=Y$\textemdash can boost performance. In another study \cite{tripathy2019privacy} using the same information-theoretical approach, a model including adversarially trained deep neural networks was presented and titled privacy-preserving adversarial networks(PPAN). The idea of adversarial training of neural networks was first introduced in \cite{goodfellow2014generative}. The PPAN model was applied to both useful data as observation $\left(W=Y\right)$ and full data as observation $\left(W=\left(X,Y\right)\right)$. The results of the PPAN model for discrete synthetic data, MNIST handwritten digits, and continuous Gaussian data were examined and found to be very close to the optimal trade-off that was derived analytically. However, the performance of the PPAN model for continuous non-Gaussian data can not be properly assessed using this framework. The reason could be this fact that generally there is no closed-form solution for the privacy-distortion trade-off. In practical applications, there are several cases (such as those related to smart meters\cite{shateri2019deep}) where attributes come from a continuous range but with the non-Gaussian distribution. Therefore, the natural challenge raised here is that how the PPAN model can be applied and evaluated in these cases? 

In this study, a method for evaluating the PPAN model for continuous non-Gaussian attributes is provided. In this method, first a lower and upper bound on the performance of the PPAN model are determined ; and then the estimated privacy-distortion of the network is compared with these bounds. These bounds include the Kraskov–Stögbauer–Grassberger (KSG) estimation of entropy and mutual information\cite{kraskov2004estimating}.

\section{Background: Privacy-Distortion Trade off Formulation}
\begin{figure}[htbp]
\centering
\includegraphics[width=0.9\linewidth]{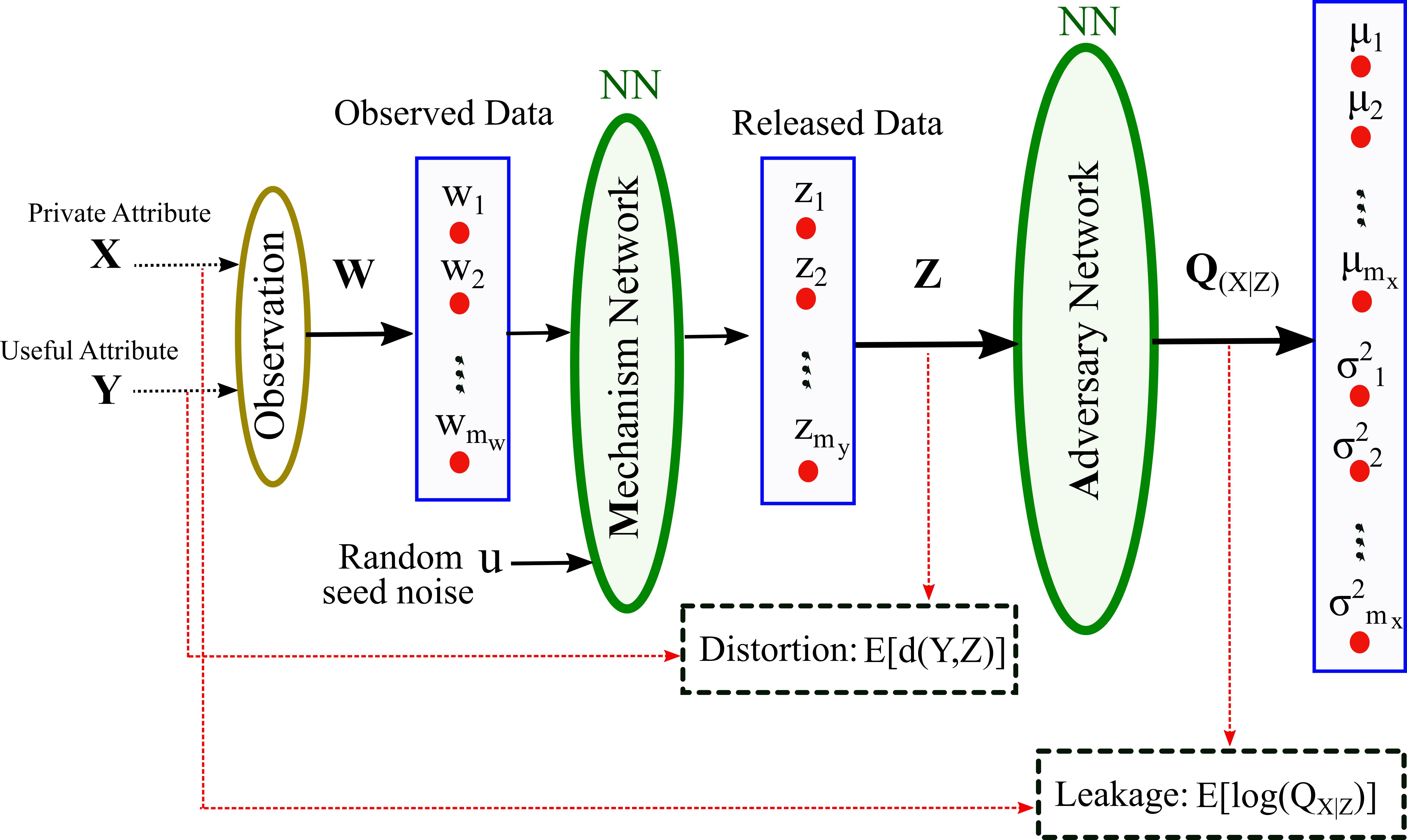}
\caption{Privacy preserving model in an adversarial fashion. Seed noise $U$ is concatenated to the observed data $W$ and is used to provide randomization for mechanism network.}
\label{had1}
\end{figure}

Consider private data $X\in \R^{m_X}$, useful data $Y\in \R^{m_Y}$, and observed data $W\in \R^{m_W}$ which are modeled as jointly distributed random variables by data model $P_{W,X,Y}$ over space $\small{\mathcal{W}\times\mathcal{X}\times\mathcal{Y}}$. The goal of the privacy-distortion trade-off model is designing a data-sharing mechanism $P_{Z|W}$ to generate released data $Z\in \R^{m_Y}$ that shares minimum information with private data $X$ while provides maximum utility of useful data $Y$. Such a mechanism can be modeled as the following optimization problem:
\begin{equation} \label{eqtr1}
\begin{aligned}
& \underset{P_{Z|W}}{\text{minimize}}
& & I(X;Z)\\
& \text{subject to}
& & \mathbb{E}\left[d\left(Y,Z\right)\right] \leq \delta
\end{aligned}
\end{equation}
\noindent where the leakage is quantified by mutual information $I\left(X;Z\right)$ between released data $Z$ and private attribute $X$ while utility is inversely quantified as expected distortion $d\left(Y,Z\right)$ between released data $Z$ and useful attribute $Y$, and parameter $\delta$ denotes the maximum allowed distortion. As it is shown in \cite{tripathy2019privacy} using variational lower bound of mutual information\cite{barber2003algorithm}, for any conditional distribution $Q_{X|Z}$ the mutual information term can be approximated as follows:
\begin{equation} \label{eqtr3}
I(X;Z) = H(X) + \underset{Q_{X|Z}}{\text{max }}\mathbb{E}\left[\log Q_{X|Z}\left(x|z\right)\right]
\end{equation}
\noindent where $H(.)$ denotes entropy and expectation $\mathbb{E}[.]$ is with respect to $P_{X|Z}(x|z)$ the true distribution over $X$ given $Z$. Substituting equation \eqref{eqtr3} in \eqref{eqtr1} and dropping constant term $H(X)$ the privacy-preserving optimization problem \eqref{eqtr1} can be written as the following minmax optimization problem:
\begin{equation} \label{eqtr4}
\underset{P_{Z|W}}{\text{min }}\underset{Q_{X|Z}}{\text{max }}\mathbb{E}\left[\log Q_{X|Z}(x|z)\right] + \lambda\left[max\left(0,\mathbb{E}\left[d\left(Y,Z\right)\right]-\delta\right)\right]^2
\end{equation}
\noindent where $\lambda>0$ is Lagrange/penalty coefficient. It should be noted that, depending on application, different distortion metrics $d(.)$ including $Pr[Y\neq Z]$ and $\mathbb{E}[\|Y-Z\|^2]$ can be used. The minimax problem \eqref{eqtr4} can be interpreted in an adversarial training context in which the adversary network uses the released data $Z$ to estimate the posterior $Q_{X|Z}(x|z)$ by maximizing the log-likelihood $\mathbb{E}[\log Q_{X|Z}(x|z)]$ while the releaser(mechanism network) attempts to prevent that by minimizing this log-likelihood (See Fig.~\ref{had1}). Following the equation \eqref{eqtr4}, the mechanism and adversary network are realized as a neural network(NN) with following loss functions:

\begin{equation} \label{eqtr5}
\begin{aligned}
& Loss_A = -\mathbb{E}\left[\log Q_{X|Z}(x|z)\right] \\
& Loss_M = \lambda \left[max\left(0,\mathbb{E}\left[d(Y,Z)\right]-\delta\right)\right]^2 - Loss_A
\end{aligned}
\end{equation}

\noindent where $Loss_A$ is the loss function of the adversary network and $Loss_M$ is the loss function of the mechanism network. 

In \cite{tripathy2019privacy} it is assumed that the posterior $Q_{X|Z}\left(x|z\right)$ is Gaussian with mean $\mu \left(Z\right)$ and covariance matrix $diag\left(\sigma^2(Z)\right)$. Applying this assumption to equation\eqref{eqtr5} and approximating expectation $\mathbb{E}(.)$ empirically by averaging over $n$ samples (considering the law of large number), equation \eqref{eqtr5} can be rewritten as follows:
\begin{equation} \label{eqtr6}
\begin{aligned}
& Loss_A \approx -\frac{1}{n}\sum_{i}\log\left[\mathcal{N}\left(x_i; \mu (z_i), \Sigma(z_i)\right)\right] \\
& Loss_M \approx \frac{1}{n}\sum_{i}\lambda\left[max\left(0,d\left(y_i,z_i\right)-\delta\right)\right]^2 - Loss_A
\end{aligned}
\end{equation}
\noindent where \normalsize \small{$\mathcal{N}(X; \mu, \Sigma) = \frac{exp\{\frac{-1}{2}(X-\mu)^T\Sigma^{-1}(X-\mu)\}}{\sqrt{2\pi \times det(\Sigma)}}$} \normalsize and $\Sigma$ is covariance matrix. The model presented in Fig.~\ref{had1} and includes two neural networks with loss functions \eqref{eqtr6} is called Privacy-Preserving Adversarial Networks (PPAN) model \cite{tripathy2019privacy}. During the training, the output of the Mechanism (releaser) and output of Adversary are used to update the parameters of each network. But after training the model, just the releaser is used.
\section{Theoretical Results}
To assess the PPAN model for cases where attributes are continuous, generally, two kinds of attributes, including Gaussian and non-Gaussian, can be considered. The continuous Gaussian case is discussed in detail and with many examples in \cite{tripathy2019privacy}. As it is mentioned in \cite{tripathy2019privacy}, for the continuous Gaussian attributes the Leakage-Distortion trade-off in equation \eqref{eqtr1} has a closed-form solution for both useful data/attribute only, $W = Y$ and full data, $W = (X,Y)$ cases. More specifically, for the useful data only (with zero mean $X$ and $Y$), and full data (with zero mean and unit variance $X$ and $Y$) cases the theoretical leakage are calculated as follows: \cite{tripathy2019privacy} 

\begin{equation} \label{eqtr7}
I_{useful}(X;Z) =max \left\{ 0, \frac{1}{2}\log \left(\frac{1}{1-\rho^{2} + \rho^2\frac{\delta}{\sigma^{2}_{Y}}}\right)\right\}
\end{equation}
\begin{equation} \label{eqtr8}
I_{full}(X;Z) = \begin{cases} 0\hspace{5.2cm} \delta \geq \rho^{2} \\ \frac{1}{2}\log \left(\frac{1}{1- \left(\sqrt{\rho^{2}(1-\delta)} - \sqrt{(1-\rho^{2})\delta}\right)^{2}}\right) \hspace{0.3cm}  \delta < \rho^{2}\end{cases}
\end{equation}

\noindent where $\rho$ is correlation coefficient X and Y and $\sigma^{2}_{Y}$ is the variance of Y. These theoretical solutions for each distortion $\delta$ are used as the baseline for the Gaussian case. In this case, after generating the released data Z using the PPAN model, the actual distortion and the leakage of the network are calculated using \eqref{eqtr9}.
\begin{equation} \label{eqtr9}
Distortion =\frac{1}{n}\sum_{i}[y_i - z_i]^2 ;\quad Leakage =-\frac{1}{2}\log (1 - \rho _{x,z}^{2})
\end{equation}
\noindent where $\rho _{x,z}$ is the correlation coefficient between private attribute X and released data Z while the leakage in equation \eqref{eqtr9} is the mutual information of bivariate Gaussian\cite{pinsker1964information}.

\subsection{Lower and upper bounds on Leakage-Distortion Trade off}
For the cases where the private and useful attributes are not Gaussian, the equations \eqref{eqtr7} and \eqref{eqtr8} for theoretical leakage calculations (baseline) and equation \eqref{eqtr9} for network leakage calculation do not hold. Therefore, in this study for non-Gaussian cases, the amount of network leakage is calculated using a non-parametric estimation of mutual information. Moreover, a lower and upper bound for the leakage-distortion tradeoff is theoretically defined using a non-parametric estimation of entropy and used as a baseline. To this end, the Kraskov–Stögbauer–Grassberger (KSG) estimation of entropy and mutual information is used. Consider a random sample of size $N$ from $d$-dimensional random vector $X = (X_1,..., X_d)$. The Kraskov–Stögbauer–Grassberger (KSG) estimation of differential entropy $\hat{H}_{knn}(X)$ based on the k-nearest-neighbor distances can be stated as follows \cite{kraskov2004estimating}:
\begin{equation} \label{eqtr11}
\hat{H}_{knn}(X) = \frac{d}{N}\sum_{i=1}^{N}\log(\epsilon_i) + \log c_d +\psi(N) - \psi(k)
\end{equation}

\noindent where $\epsilon_i$ is twice the distance from $x_i$ to its $k^{-th}$ neighbor, $\psi(m) = \frac{d}{dm}\ln(\Gamma (m))$ is digamma function, and $c_d$ is the volume of a unit $d$-dimensional ball which for the Euclidean norm $c_d = \frac{\pi^{d/2}}{2^d\Gamma(d/2+1)}$. Considering equation \eqref{eqtr11} and the expansion of mutual information as $I(X;Z) = H(X) + H(Z) - H(X,Z)$, the KSG estimation of $I(X;Z)$ is calculated as follows\cite{kraskov2004estimating}:
\begin{equation} \label{eqtr12}
\hat{I}_{knn}(X;Z) = \psi(N) + \psi(k) - <\psi(n_x +1) + \psi(n_z +1)>
\end{equation}

\noindent where $<.>$ denotes to averages both over $i\in [1, ..., N]$ and over all realization of random samples, $n_x(i)$ is the number of points $x_j$ whose distance from $x_i$ is less than $\epsilon(i)/2$, $n_z(i)$ is the number of points $z_j$ whose distance from $z_i$ is less than $\epsilon(i)/2$, and $\epsilon(i) = max\{\epsilon_x(i), \epsilon_z(i)\}$.\\

In this work, for the non-Gaussian attributes, equations \eqref{eqtr11} and \eqref{eqtr12} are used to assess the PPAN model performance. To this end, two different kinds of observation including useful data only, $W = (Y,U)$ and full data, $W = (X,Y,U)$ are discussed separately at the rest of this study. It should be noted that all the findings are for the case where attributes have zero mean.\\

\subsubsection{Useful data only} When the private and useful attributes are not Gaussian, there is no closed form solution for the optimization problem \eqref{eqtr1}. In this case by considering 
\begin{equation*}
 \mathbb{R}(\delta) = \underset{W=Y, \mathbb{E}[(Y-Z)^2]\leq \delta}{\text{min}}I(X;Z)  
\end{equation*}
 
to determine a lower bound for $\mathbb{R}(\delta)$ we can say:
\begin{equation} \label{eqtr13}
\begin{split}
I(X;Z)&=H(X) - H(X|Z)= H(X) - H(X - \rho\sigma_X Z/\sigma_Y|Z)\\
&\geq H(X)- H(X - \rho\sigma_X Z/\sigma_Y)\\
&\geq H(X) - H(\mathcal{N}(0,\mathbb{E}[(X - \rho\sigma_X Z/\sigma_Y)^2]))\\
&= H(X) - 0.5\log\left(2\pi e \mathbb{E}\left[(X - \rho\sigma_X Z/\sigma_Y)^2\right]\right)\\
\end{split}
\end{equation}
\noindent where the first inequality in \eqref{eqtr13} holds because conditioning reduces entropy and the second inequality is true since for a given value of second moment, the zero-mean normal distribution has maximum entropy. The first term in inequality \eqref{eqtr13} is replaced by entropy estimation \eqref{eqtr11}. According to\cite{tripathy2019privacy} by considering distortion condition $\mathbb{E}[(Y-Z)^2]\leq \delta$, it can be said that: $\mathbb{E}[(X - \rho\sigma_X Z/\sigma_Y)^2]\leq \sigma_X^2(1-\rho^2) + \rho^2\delta\sigma_X^2/\sigma_Y^2$. Therefore, by considering this fact that mutual information is non-negative, we have:\normalsize
 
\begin{dmath} \label{eqtr14}
\mathbb{R}(\delta) \geq max\left\{0,\hat{H}_{knn}(X) -0.5\log\left[2\pi e\sigma_X^2\left(\rho^2\frac{\delta}{\sigma_Y^2} +(1-\rho^2)\right)\right]\right\}
\end{dmath}
\normalsize
To find an upper bound for $\mathbb{R}(\delta)$, following the procedure in \cite{rebollo2010t}, and considering the convexity of mutual information and mutual information estimation\eqref{eqtr12} we have:
\begin{equation} \label{eqtr15}
\mathbb{R}(\delta)\leq \hat{I}_{knn}(X;Y)\times \small{(1 - \frac{\delta}{\sigma_Y^2})}
\end{equation}
Therefore, equations \eqref{eqtr14} and \eqref{eqtr15} together form the lower and upper bounds on the performance of the PPAN model for the non-Gaussian attributes with useful data as the observation. For unit variance attributes, at $\delta = 1$ the lower and upper bounds intersect. While for Gaussian attributes they intersect at $\delta = 0$.\\

\subsubsection{Full data} For the non-Gaussian attributes, similar to the previous case, there is no closed-form solution for the optimization problem \eqref{eqtr1}. The upper bound in this case is similar to \eqref{eqtr15} but the lower bound for the unit variance attributes would be different. Following the procedure in \cite{tripathy2019privacy}, by considering the linear minimum mean squared error estimation of $X$ given $Z$ as \small{$\hat{\mathbb{E}}[X|Z] = \mathbb{E}[XZ]Z/\mathbb{E}[Z^2]$} \normalsize  and orthogonality rule in least squares estimation \small{$\mathbb{E}\left[\hat{\mathbb{E}}[X|Z](X-\hat{\mathbb{E}}[X|Z]) \right] = 0$} \normalsize  we have:
\begin{equation} \label{eqtr16}
\begin{split}
I(X;Z)&=H(X) - H(X|Z) = H(X) - H\left(X - \hat{\mathbb{E}}[X|Z]\bigg|Z\right)\\
&\geq H(X)- H\left(X - \hat{\mathbb{E}}[X|Z]\right)\\
&\geq H(X) - H\left(\mathcal{N}\left(0,\mathbb{E}[(X - \hat{\mathbb{E}}[X|Z])^2]\right)\right)\\
&= H(X) - 0.5\log\left(2\pi e\left(\mathbb{E}[X^2]-\frac{\mathbb{E}^2[XZ]}{\mathbb{E}[Z^2]}\right) \right)\\
\end{split}
\end{equation}
\noindent where similar to \eqref{eqtr13}, the first inequality in \eqref{eqtr16} holds because conditioning reduces entropy and the second inequality is true since, for a given value of the second moment, the zero-mean normal distribution has maximum entropy. Now to find the lower bound of $\mathbb{R}(\delta)$, the minimum of term $\mathbb{E}^2[XZ]/\mathbb{E}[Z^2]$ should be found by considering the distortion condition. To this end, for unit variance attributes, by considering random variables $X,Y,Z$ as vectors $\textbf{x}$,$\textbf{y}$,$\textbf{z}$ in vector space $\ell_2$ (and therefore replacing expectation operator on the product of two random variables with inner product in $\ell_2$) we have\cite{tripathy2019privacy}:
\begin{equation}\label{eqtr17}
 \underset{\mathbb{E}[(Y-Z)^2]\leq \delta}{\text{min}}\frac{\mathbb{E}^2[XZ]}{\mathbb{E}[Z^2]} =  \underset{\|\textbf{y}-\textbf{z}\|^2\leq \delta}{\text{min}}\frac{|<\textbf{x},\textbf{z}>|^2}{\|\textbf{z}\|^2}
 \end{equation}
By associating $\hat{i}:=\textbf{x}$, $\small{\hat{j}:=\frac{1}{\sqrt{1-\rho^2}}(\textbf{y}-\rho\textbf{x})}$, and $\hat{k}:=\frac{\textbf{z}-Proj_{Span(\textbf{x},\textbf{y})}}{\|\textbf{z}-Proj_{Span(\textbf{x},\textbf{y})}\|}$ as the unit vectors along orthogonal coordinate axes, and by considering $\textbf{t}:=\textbf{z}-\textbf{y} = t_1\hat{i} + t_2\hat{j} + t_3\hat{k}$, it can be said that $\textbf{x}=\hat{i}$, $\small{\textbf{y}=\rho\hat{i} + \sqrt{1-\rho^2}\hat{j}}$, and $\small{\textbf{z}=(t_1+\rho)\hat{i}+(t_2+\sqrt{1-\rho^2})\hat{j}+t_3\hat{k}}$. Therefore, by substituting $\textbf{x}$,$\textbf{y}$,$\textbf{z}$ in the previous minimization problem, it changes as the following optimization problem: 
\begin{equation}\label{eqtr18}
\underset{t_1^2+t_2^2+t_3^2\leq \delta}{\text{min}}\left[\frac{(t_1+\rho)^2}{t_1^2+t_2^2+t_3^2+2t_1\rho+2t_2\sqrt{1-\rho^2}+1}\right]
\end{equation}

It is shown in \cite{tripathy2019privacy} that when $\rho^2 \leq \delta$, problem \eqref{eqtr18} is minimum for $t_1^*=-\rho, t_2^*=t_3^*=0$ while for the case $\rho^2 > \delta$ the minimum of \eqref{eqtr18} is attained for $\small{t_1^*=-\delta\rho-\sqrt{\delta(1-\delta)(1-\rho^2)}, t_2^*=\sqrt{\delta-(t_1^*)^2}, t_3^*=0}$. By substituting the $t_1^*, t_2^*, t_3^*$ in \eqref{eqtr18} we have:
\begin{equation}\label{eqtr19}
\underset{\mathbb{E}[(Y-Z)^2]\leq \delta}{\text{min}}\frac{\mathbb{E}^2[XZ]}{\mathbb{E}[Z^2]} =\small{\begin{cases} 0\hspace{3.8cm} \delta \geq \rho^{2} \\ (\sqrt{\rho^{2}(1-\delta)} - \sqrt{(1-\rho^{2})\delta})^{2} \hspace{0.4cm}  \delta < \rho^{2}\end{cases}}
\end{equation}
Therefore, by considering \eqref{eqtr17} and \eqref{eqtr19} and entropy estimation $\hat{H}_{knn}(X)$ and $\mathbb{E}[X^2]=1$(zero mean unit variance attributes) the lower bounds of $\mathbb{R}(\delta)$ for full data case with unit variance attributes $\mathbb{R}_{UVLB}(\delta)$ (unit-variance lower bound) is as follows:
\begin{equation}\label{eqtr20}
\begin{split}
\mathbb{R}_{UVLB}(\delta) =&\small{\begin{cases}c\hspace{5.8cm} \delta \geq \rho^{2} \\ c-0.5\log\left[1- \left(\sqrt{\rho^{2}(1-\delta)} - \sqrt{(1-\rho^{2})\delta}\right)^{2}\right] \hspace{0.25cm}  \delta < \rho^{2}\end{cases}}
\end{split}
\end{equation}
\noindent where $c = \hat{H}_{knn}(X) - 0.5\log(2\pi e)$. Since the mutual information is non-negative we have:
\begin{equation}\label{eqtr21}
\mathbb{R}(\delta) \geq max\left\{0, \mathbb{R}_{UVLB}(\delta)\right\}
\end{equation}

For the non-unit variance attributes, generally, the lower bounds would be very complicated, however for the case $\sigma^2_X = \sigma^2_Y = \sigma^2$ the approach for finding the lower bound is exactly similar until arriving to equation \eqref{eqtr17}. Now, to solve equation \eqref{eqtr17} (for $\sigma^2_X = \sigma^2_Y = \sigma^2$), by associating $\hat{i}:=\textbf{x}/\sigma$, $\small{\hat{j}:=\frac{1}{\sigma\sqrt{1-\rho^2}}(\textbf{y}-\rho\textbf{x})}$, and $\hat{k}:=\frac{\textbf{z}-Proj_{Span(\textbf{x},\textbf{y})}}{\|\textbf{z}-Proj_{Span(\textbf{x},\textbf{y})}\|}$ as the unit vectors along orthogonal coordinate axes and by considering $\textbf{t}:=\textbf{z}-\textbf{y} = t_1\hat{i} + t_2\hat{j} + t_3\hat{k}$, we can say $\small{\textbf{x}=\sigma\hat{i}}$, $\small{\textbf{y}=\sigma\rho\hat{i} + \sigma\sqrt{1-\rho^2}\hat{j}}$, and $\small{\textbf{z}=(t_1+\sigma\rho)\hat{i}+(t_2+\sigma\sqrt{1-\rho^2})\hat{j}+t_3\hat{k}}$. By substituting these $\textbf{x}$, $\textbf{y}$, and $\textbf{z}$ in \eqref{eqtr17}, the minimization problem \eqref{eqtr17} changes as follows:
\begin{equation}\label{eqtr22}
\underset{t_1^2+t_2^2+t_3^2\leq \delta}{\text{min}}\left[\frac{\sigma^2(t_1+\sigma\rho)^2}{t_1^2+t_2^2+t_3^2+2t_1\sigma\rho+2t_2\sigma\sqrt{1-\rho^2}+\sigma^2}\right]
\end{equation}

By factoring $\sigma^2$ from numerator and denominator of the objective function in \eqref{eqtr22} and the both sides of the distortion constraint, and then using change of variables as $t\textprime_1=\frac{t_1}{\sigma}$, $t\textprime_2=\frac{t_2}{\sigma}$, $t\textprime_3=\frac{t_3}{\sigma}$ and considering $\delta\textprime = \delta/\sigma^2$ the minimization problem \eqref{eqtr22} changes as follows:
\begin{equation}\label{eqtr24}
\underset{t\textprime_1^2+t\textprime_2^2+t\textprime_3^2\leq \delta\textprime}{\text{min}}\left[\frac{\sigma^2(t\textprime_1+\rho)^2}{t\textprime_1^2+t\textprime_2^2+t\textprime_3^2+2t\textprime_1\rho+2t\textprime_2\sqrt{1-\rho^2}+1}\right]
\end{equation}

\noindent which is exactly similar to the minimization problem \eqref{eqtr18} except with a $\sigma^2$ factor at the numerator of the objective function. Therefore, similar to the \eqref{eqtr18}, when $\rho^2 \leq \delta\textprime$, problem \eqref{eqtr24} is minimum for $t\textprime_1^*=-\rho, t\textprime_2^*=t\textprime_3^*=0$ while for the case when $\rho^2 > \delta\textprime$ the minimum of \eqref{eqtr24} is found for $\small{t\textprime_1^*=-\delta\textprime\rho-\sqrt{\delta\textprime(1-\delta\textprime)(1-\rho^2)}, t_2^*=\sqrt{\delta\textprime-(t\textprime_1^*)^2}, t\textprime_3^*=0}$. Thus, for the non-unit variance and when $\sigma^2_X = \sigma^2_Y = \sigma^2$ we have:
\begin{equation}\label{eqtr25}
\begin{split}
\underset{\mathbb{E}[(Y-Z)^2]\leq \delta}{\text{min}}\frac{\mathbb{E}^2[XZ]}{\mathbb{E}[Z^2]} =&\small{\begin{cases} 0\hspace{4.35cm} \delta\textprime \geq \rho^{2} \\ \sigma^2\left(\sqrt{\rho^{2}(1-\delta\textprime)} - \sqrt{(1-\rho^{2})\delta\textprime}\right)^{2} \hspace{0.25cm}  \delta\textprime < \rho^{2}\end{cases}}\\
=& \small{\begin{cases} 0\hspace{4cm} \delta \geq \sigma^2\rho^{2} \\ \left(\sqrt{\rho^{2}(\sigma^2-\delta)} - \sqrt{(1-\rho^{2})\delta}\right)^{2} \hspace{0.25cm}  \delta < \sigma^2\rho^{2}\end{cases}}
\end{split}
\end{equation}
Therefore, by considering \eqref{eqtr16} and \eqref{eqtr25} and entropy estimation $\hat{H}_{knn}(X)$ and $\mathbb{E}[X^2]=\sigma^2_X =\sigma^2$, the lower bounds of $\mathbb{R}(\delta)$ for full data case and non-Gaussian attributes with $\sigma^2_X = \sigma^2_Y = \sigma^2$ is as follows:
\begin{equation}\label{eqtr26}
\begin{split}
\mathbb{R}_{LB}(\delta) = &\small{\begin{cases} c- 0.5\log(\sigma^2)\hspace{4.3cm} \delta \geq \sigma^2\rho^{2} \\ c- 0.5\log\left[\sigma^2- \left(\sqrt{\rho^{2}(\sigma^2-\delta)} - \sqrt{(1-\rho^{2})\delta}\right)^{2}\right] \hspace{0.16cm}  \delta < \sigma^2\rho^{2}\end{cases}}
\end{split}
\end{equation}
Similarly, since mutual information is non-negative we have:
\begin{equation}\label{eqtr27}
\mathbb{R}(\delta) \geq max\left\{0, \mathbb{R}_{LB}(\delta)\right\}
\end{equation}

\section{Examples on performance of PPAN model}
In this section, the PPAN model is evaluated for several cases where useful data $Y$ and private data $X$ are continuous non-Gaussian. In all the examples, both the Mechanism and Adversary networks are deep neural network each include two hidden layers with 16 nodes and rectified linear unit (ReLu) as the activation function. The data set includes 8000 training and 4000 test samples where 10\% of the training is used as validation data for tuning hyperparameters including minibatch size $B$, penalty coefficient $\lambda$, number of steps applied for training the adversary $\ell$, and the width of seed noise $m_{noise}$. In the following examples, the seed noise $U$ is generated from independent and identically distributed (i.i.d.) samples according to a uniform distribution on the interval$[0,1]$.

\subsection{Continuous non-Gaussian Synthetic Data Set}

\begin{figure}[htbp]
\centering
\includegraphics[width=0.9\linewidth]{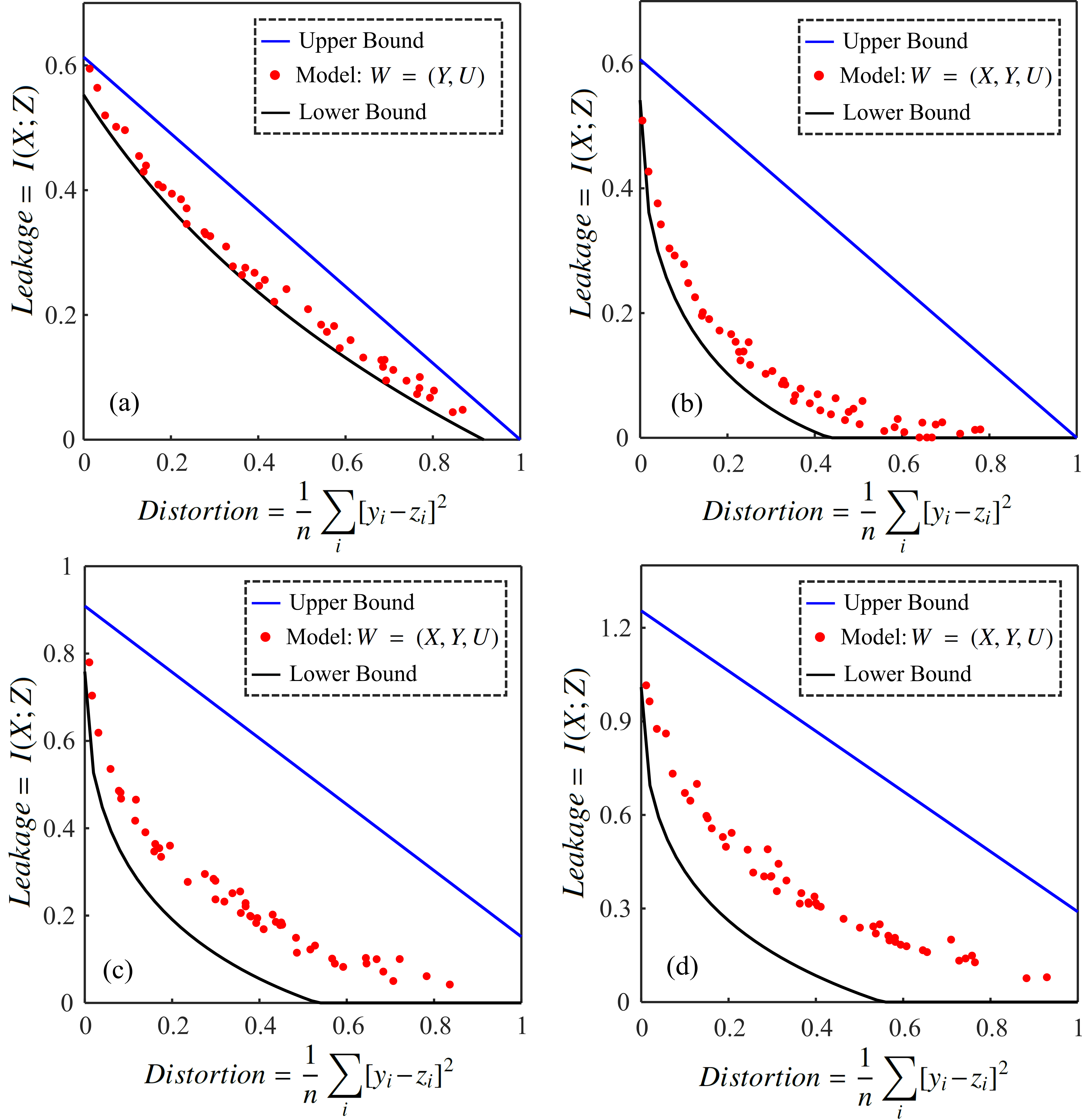}
\caption{The results of privacy-distortion trade of for the test data set. (a) Gaussian mixture for the $W = Y$ where $ \left(B,\lambda,\ell,m_{noise}\right)= \left(200,2,4,7\right)$. (b) Gaussian mixture for the $W = (Y, X)$ where $ \left(B,\lambda,\ell,m_{noise}\right)= \left(200,10,2,3\right)$. (c) Multivariate Laplace with $ \left(B,\lambda,\ell,m_{noise}\right)= \left(200,10,4,5\right)$. (d) Multivariate Uniform on the interval $[0,1]$ with $ \left(B,\lambda,\ell,m_{noise}\right)= \left(200,10,2,5\right)$.}
\label{had3}
\end{figure}

For this part, three examples are provided where useful data $Y$ and private data $X$ come from Continuous non-Gaussian Synthetic Data. As the first example, assume useful data $Y$ and private data $X$ each come from a Gaussian mixture of three random variables with 4000 samples from each distribution as: $\small{\begin{bmatrix} X_1\\Y_1 \end{bmatrix} \sim \mathcal{N}\left(\begin{bmatrix} 0\\0 \end{bmatrix},\begin{bmatrix} 1&0.85\\0.85&1 \end{bmatrix}\right)}$, $\small{\begin{bmatrix} X_2\\Y_2 \end{bmatrix} \sim \mathcal{N}\left(\begin{bmatrix} 1\\1 \end{bmatrix},\begin{bmatrix} 1.5&0.95\\0.95&1.5 \end{bmatrix}\right)}$, and $\small{\begin{bmatrix} X_3\\Y_3 \end{bmatrix} \sim \mathcal{N}\left(\begin{bmatrix} -1\\-1 \end{bmatrix},\begin{bmatrix} 0.5&0.35\\0.35&0.5 \end{bmatrix}\right)}$. This example is considered for two cases including observation based on the useful data, and observation based on the full data. As the second example, the private and useful attributes $X$ and $Y$ are multivariate Laplace distribution with covariance matrix $\begin{bmatrix} 1.2&0.90\\0.90&1.2 \end{bmatrix}$. Finally, in the third example the private and useful attributes $X$ and $Y$ are multivariate Uniform on the interval $[0,1]$ with covariance matrix $\small{\begin{bmatrix} 1.3&0.95\\0.95&1.3 \end{bmatrix}}$. The last two examples are considered for the case where observation is based on the full data. The results of applying the PPAN model to these examples are presented in Fig.~\ref{had3}. From this figure, it can be seen that the results of PPAN are within the determined bounds close to the lower bound.

\subsection{Continuous non-Gaussian Real Data Set}

\begin{figure}[htbp]
\centering
\includegraphics[width=0.55\linewidth]{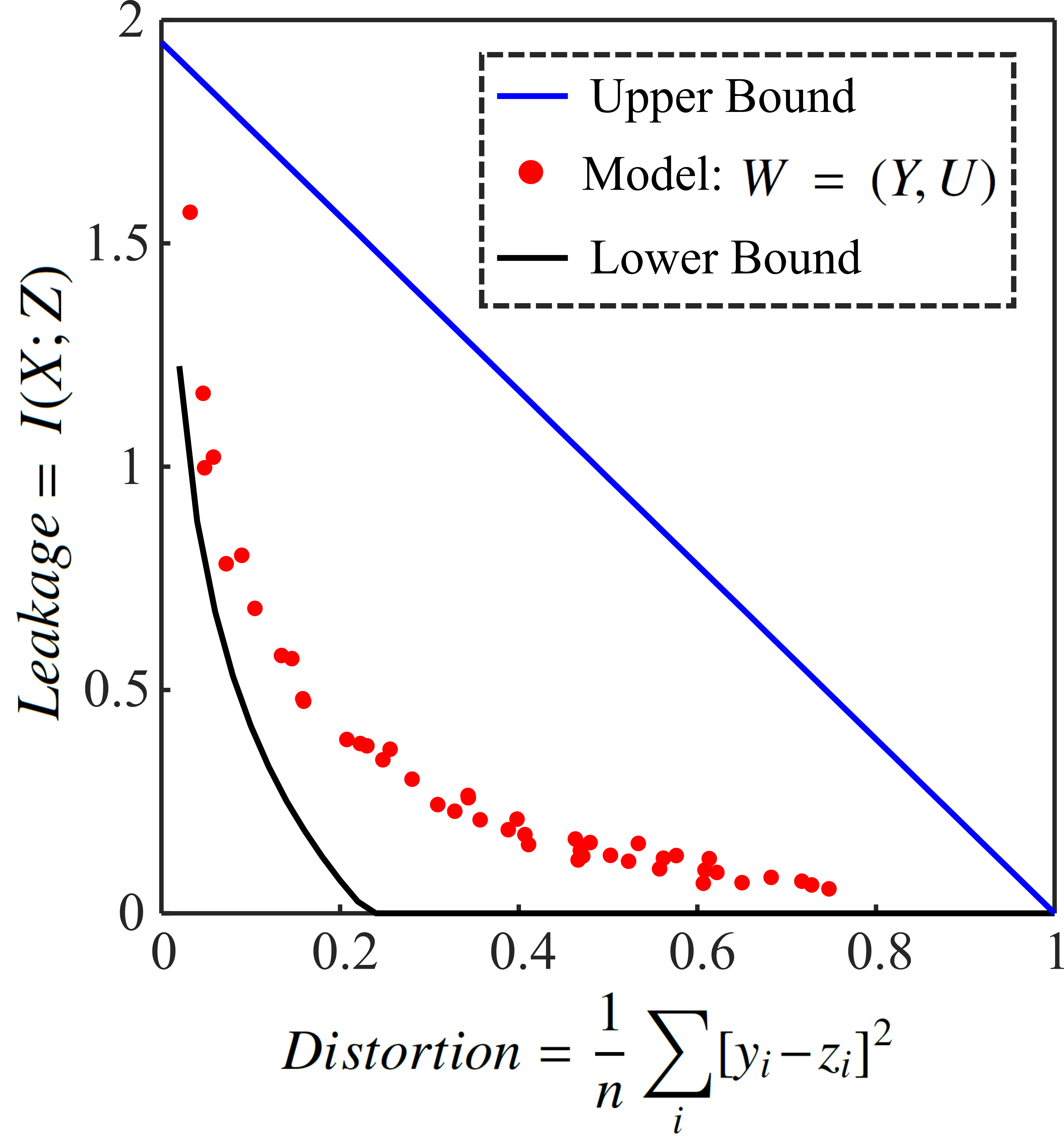}
\caption{Test results of privacy-distortion trade of for the Pecan Street data set where $ \left(B,\lambda,\ell,m_{noise}\right)= \left(200,10,3,5\right)$.}
\label{had4}
\end{figure}

In this part, the PPAN model is applied to a practical case where we aim at sharing electricity consumption of several houses in a certain region with a third party and at the same time prevent any adversary to estimate the actual pattern of household consumption out of the shared data. To this end, the Pecan Street data set  which contains hourly electricity consumption of houses in Texas, Austin is used \cite{street2019dataport}. In this example, useful data $Y$ and private data $X$ both are the same as electricity consumption. The result of using the PPAN model, in this case, is presented in Fig.~\ref{had4}. 

From the provided examples it can be concluded that similar to the Gaussian case, the PPAN model can be used for continuous non-Gaussian attributes as well, but with this difference that the assessment of the model is done using mutual information estimation compared with the provided lower and upper bounds.
 
\section{Conclusion}
In this study, the privacy-preserving adversarial network (PPAN) was examined for continuous non-Gaussian attributes. Although the PPAN model showed worked well for the discrete synthetic data and continuous Gaussian attributes, no method was provided for assessing this model for continuous non-Gaussian attributes. In this work, the performance of the PPAN model was assessed using the KSG estimation of entropy and mutual information. A lower bound and an upper bound of the privacy-distortion problem was determined using entropy and mutual information estimation and the result of the PPAN was compared with these two bounds. Several examples based on synthetic data set are provided. In addition, a practical case related to sharing power consumption of household (using actual dataset) were examined.The results showed that the performance of the PPAN for these cases is within the determined bounds close to the lower bound.


\bibliographystyle{ieeetr}
\bibliography{Main}

\vspace{12pt}

\end{document}